\documentstyle[12pt]{article}
\newcommand\beq{\begin{equation}}
\newcommand\eeq{\end{equation}}
\newcommand\bea{\begin{eqnarray}}
\newcommand\eea{\end{eqnarray}}

\begin{document}
\vspace{-2.0cm}
\bigskip

\begin{center} 
{\Large \bf  Coherent States with SU(N) Charges} 
\end{center} 
\vskip .8 true cm

\begin{center} 
{\bf Manu Mathur} \footnote{manu@bose.res.in} and 
{\bf Samir K. Paul}\footnote{smr@bose.res.in} 
\vskip 0.5 true cm

S. N. Bose National Centre for Basic Sciences \\ 
JD Block, Sector III, Salt Lake City, Calcutta 700 098, India

\vskip .1 true cm

\end{center} 
\bigskip

\centerline{\bf Abstract}

\noindent We define coherent states carrying SU(N) charges  
by exploiting generalized Schwinger boson representation of SU(N) Lie 
algebra. These coherent states are defined on $2~(2^{N - 1} - 1)$ 
complex planes. They satisfy continuity property and provide 
resolution of identity.  We also exploit this technique to construct 
the corresponding non-linear SU(N) coherent states. 

\vskip .4 true cm
\noindent PACS: ~02.20.-a 
\vskip .4 true cm

\section{\bf Introduction}

The simplest coherent states are associated with the Heisenberg-Weyl 
group with the Lie algebra involving a harmonic oscillator  
creation-annihilation operators $(a^{\dagger}, a)$ obeying: 
\bea
[a,a^{\dagger}] = {\cal I}. 
\label{hw} 
\eea 
\noindent In (\ref{hw}), ${\cal I}$  is the identity operator.  
The coherent states belonging to the Heisenberg-Weyl group 
are defined as \cite{schh, glauber, sudarshan, klauder}: 
\bea 
\vert z \rangle_{_{\infty}}  \equiv  exp ({z a^{\dagger}}) \vert 0 
\rangle  = \sum_{n=0}^{\infty} \frac {z^{n}}{\sqrt{n!}} \vert n>  
\label{cs1} 
\eea 
The subscript $\infty$ in (\ref{cs1}) denotes that the coherent states 
belong to the infinite dimensional representation of the Heisenberg-Weyl 
Lie algebra (\ref{hw}).  The manifold corresponding to the Heisenberg-Weyl 
group is the complex z plane and the corresponding coherent states 
(\ref{cs1}) are continuous functions of z. Further, they are eigenstates 
of the annihilation operator: 
\bea 
a\vert z \rangle_{\infty}  = z ~ \vert z \rangle_{\infty} . 
\label{es} 
\eea 
It is also easy to verify  that $\vert z \rangle_{\infty}$ satisfy 
the resolution of identity: 
\bea 
\int d\mu(z) \vert z \rangle_{_{\infty}}~ {}_{_{\infty}}\langle z \vert 
\equiv {\cal I}.  
\label{roihw}  
\eea
\noindent In (\ref{roihw}), $d\mu{(z)} \equiv exp\{ -(|{z}|)^{2} \}$ is the  
measure over the group manifold. The coherent states in (\ref{cs1}) can be 
generalized to define 
non-linear coherent states by demanding: 
\bea 
f(N)~a \vert z, f \rangle_{\infty} = z~\vert z, f \rangle_{\infty} 
\label{nlhw}
\eea 
where $f(\hat{N})$ is an operator valued function of the number operator 
$\hat{N} = a^{\dagger}a$. 
An explicit expression for $\vert z, f \rangle_{\infty}$ is given by: 
\bea 
\vert z, f \rangle_{\infty} = \sum_{n=0}^{\infty} \frac{c(n)}
{\sqrt{n!}}~z^{n}~|n>  
\label{nlhwe} 
\eea 
where, c(0) = 1, and $c(n) = \prod_{k=0}^{n-1} [f(k)]^{-1}$  
if $n \ge  1$.  The expansion in (\ref{nlhwe}) can be summed 
over to get an exponential form: 
\bea 
\vert z, f \rangle_{\infty} = exp \Big(z \hat{H}_f(\hat{N}) a^{\dagger} 
\Big)~|0>, 
\label{nlhwe2}
\eea
where $ \hat{H}_f(\hat{N}) \equiv {1 \over f(\hat{N}-1)}$.  
Note that $\vert z, f = 1 \rangle_{\infty} = \vert z \rangle_{\infty}$. 

\noindent In this paper, we generalize the above simple ideas associated 
with the Heisenberg-Weyl group to SU(N) group. For this we need to 
generalize the Schwinger boson representation of SU(2) Lie 
algebra with a doublet of harmonic oscillators \cite{schwinger} to 
SU(N) group.  This representation of the SU(N) Lie algebra in terms of 
certain number of harmonic oscillators will help us define coherent states  
which are characterized by (N-1) charges. These (N - 1) charges are the 
eigenvalues of the (N-1) generators belonging to the Cartan subspace of 
SU(N). The coherent states with the above property  will be 
called fixed charge coherent states or more explicitly, 
coherent states with SU(N) charges. 
The fixed charge coherent states also satisfy certain eigenstate 
conditions which are similar to (\ref{es}) and thus 
are very different from the `generalized coherent states' defined by 
Perelomov \cite{perelomov}.  More specifically, while the previous 
ones, by definition,  are the eigenstates of (N - 1) diagonal generators 
belonging to the Cartan subspace of SU(N), the generalized SU(N) 
coherent states  are eigenstates of the (N - 1) Casimirs of SU(N), 
i.e, they are defined within various irreducible representations of SU(N). 
Further, the fixed charge coherent states are defined on 
complex planes while the 
later are defined on the SU(N) group manifolds which are compact. 
Infact, this work 
corresponds to a simple generalization and a co-variant formulation  
of the work done in \cite{bdr} for SU(2) group to SU(N) group for 
arbitrary N. 
The techniques using the generalized Schwinger representation of SU(N) Lie 
algebra \cite{mm3, mmn} developed in this paper also lead to a unified 
formulation of  the above two types 
of coherent states. This will be clear as we proceed. 
We further generalize this construction of fixed charge coherent states 
to define non-linear fixed charge SU(N) coherent states. This  corresponds to 
generalizing the well studied non-linear coherent states in (\ref{nlhwe2}) 
of the Heisenberg-Weyl group \cite{nl1,nl2,nl3,nl4} to SU(N) group. 
At this stage we should mention that the work in \cite{chin} 
deals with SU(3) charged coherent states and many ideas of this paper are 
borrowed and generalized to SU(N) from there. However, the work \cite{chin} 
is not  complete as it involves only symmetric 
representations of SU(3) and therefore does not reveal the much richer 
contents of the SU(3) charged coherent states (section 3). 

\noindent The plan of the paper is as follows: In section 2 , we deal with 
the SU(2) group and define SU(2) charge as well as spin  or generalized 
SU(2) coherent states using the Schwinger boson representation.  This 
formulation will put the above two types of coherent states on  similar 
footing.  We then deal with SU(2) non-linear charge coherent states in 
section 2.1 . The section 2 contains some of the results of \cite{bdr} 
in a co-variant language which is necessary for it's generalization to 
SU(N) groups. It, therefore, provides a platform to understand 
the SU(N) ($N \ge 3$) techniques of section 3 and section 4 
within the much simpler SU(2) framework. The section 3 
deals with SU(3) group and defines SU(3) charge coherent states and 
their non-linear analogs. 
This section is required for the continuity before going to SU(N) for 
arbitrary N as SU(2) group is too simple. More explicitly, the SU(N) 
groups for $N \ge 3$ has more than one fundamental representations which 
requires a much more careful analysis. This will be emphasized in 
section 3 and section 4. Thus, in section 3 we find many features/results 
for SU(3) which were redundant/absent for SU(2) group. 
Finally, in section 4 we straight-away generalize the contents 
of section 3 to SU(N) for arbitrary N. 
   
\section{\bf Coherent States with SU(2) Charge}

We start this section with the standard definition of generalized SU(2) 
coherent states \cite{perelomov}. We then introduce Schwinger bosons 
\cite{schwinger} to construct SU(2) Lie algebra and exploit it to give 
an easier formulation to construct the above coherent states. This 
formulation with some simple changes leads to fixed charge SU(2) 
coherent states.  
The SU(2) group involves 3 angular momentum generators, $J_1, J_2$ and 
$J_3$ and the Lie algebra is: 

\bea
[J^{a},J^{b}] = i~\epsilon_{abc}~ J^{c}, ~~~ a,b,c =1,2,3.   
\label{am}  
\eea 

\noindent The SU(2) Casimir is $\vec{J}.\vec{J}$ with eigenvalue 
j(j+1) where j is half integer spin. To construct the SU(2) spin 
coherent states,  we define the group element 
U using the three Euler angles, i.e., $U(\theta,\phi,\psi) \equiv \exp 
(i \phi J_3) \exp (i \theta J_2) \exp (i \psi J_3)$.  Further, in the  
spin j representation, we use the highest weight $\vert j,j \rangle$ 
as the chosen representation state vector.  The SU(2) coherent states 
are defined as: 
\bea
\vert \theta,\phi,\psi) >_{j} ~&=&~ U(\theta,\phi,\psi) ~\vert j, j> ~,
\nonumber \\
&=&~ exp  (i~j\psi) \sum_{m=-j}^{+j} C_{m}(\theta,\phi) ~\vert j, m> ~,
\label{st}
\eea 

\noindent In (\ref{st}), 
\bea 
C_{m}(\theta,\phi) = \sqrt{\frac{2j!}{(j+m)!(j-m)!}}~ exp(im\phi)~ 
\Big(sin \frac{\theta}{2}\Big)^{j-m}~ 
\Big(cos \frac{\theta}{2}\Big)^{j+m}
\label{cof} 
\eea

\noindent The above construction is simple. However, for larger Lie 
group and in particular for SU(N) $N \ge 3$, it becomes extremely 
difficult to compute the expansion coefficients 
analogous to $C_{m}(\theta,\phi)$. This problem is completely evaded by 
another equivalent formulation which exploits the  Schwinger boson 
representation of the corresponding Lie algebra \cite{mm3,mmn}. Further, 
this method like Heisenberg construction of section 1,  
involves harmonic oscillator creation anihilation operators and resembles 
(\ref{cs1}). We will illustrate this procedure for SU(2) in this section 
before going to SU(N) for arbitrary N. 

\noindent The algebra in (\ref{am}) can be realized in terms of
a doublet of harmonic oscillator creation and annihilation operators
$\vec{a} \equiv (a_1, a_2)$ and $ \vec{a}^{\dagger} \equiv (a^{\dagger}_1,
a^{\dagger}_2)$ respectively \cite{schwinger}. The number operators are 
$\hat{N}_1 \equiv a^{\dagger}_{1}a_{1}$ and $\hat{N}_2 \equiv 
a^{\dagger}_{2}a_{2}$. They satisfy the bosonic commutation relations: 
\bea
[a_i,a^{\dagger}_j] = \delta_{ij},~~~  i,j =1,2. 
\label{ho} 
\eea 
\noindent The vacuum state is $\vert 0,0 >$ and an arbitrary state 
in the number operator basis is written as $\vert n_1,n_2 >$. It 
satisfies: 
\bea
\hat{N}_1 ~ \vert n_1,n_2 >  =  n_1~\vert n_1,n_2 >, ~~~ 
\hat{N}_2 ~ \vert n_1,n_2 >   =  n_2~ \vert n_1,n_2 >. 
\label{no}
\eea 

\noindent We can now define the angular momentum operators in (\ref{am}) 
as: 
\beq
J^a ~\equiv  {1 \over 2} ~a^{\dagger}_i ~(\sigma^a)_{ij} ~a_j ~,
\label{sch}
\eeq
\noindent where $\sigma^a$ denote the Pauli matrices.  
It is easy to check that the operators in (\ref{sch}) satisfy (\ref{am}). 
Further, as they involve one creation and one annihilation operator, 
the Casimir $({\cal C})$ is: 
\bea 
{\cal C}  = \hat{N}_1  +   \hat{N}_2.  
\label{cas} 
\eea 
\noindent One can also explicitly check that $\vec{J} \cdot \vec{J} 
\equiv {1 \over 4} {\cal C} ({\cal C} + 2) =
{1 \over 4} \vec{a}^{\dagger} \cdot \vec{a}
(\vec{a}^{\dagger} \cdot \vec{a} + 2)$. Thus the representations of $SU(2)$
can be characterized by the eigenvalues of the total occupation number 
operator and the spin value $j$ is equal to $n/2 \equiv (n_1+ n_2)/2$.  
With the SU(2) Schwinger representation (\ref{sch}), we can directly 
generalize (\ref{cs1}) to SU(2)\footnote{A similar construction using 
Holstein Primakov representation of SU(2) Lie algebra was given in 
\cite{radcliffe}. We find that (\ref{cs2}) is simpler and easily 
generalizable to SU(N).}:

\bea
\vert \vec{z}>_{n=2j} & = & \delta_{[{a_1}^{\dagger}{a_1}
+{a_2}^{\dagger} a_2,n]}
~ \exp \Big( \vec{z} \cdot \vec{a}^\dagger \Big) ~\vert 0,0> \nonumber \\ 
& = & n!~\sum_{r=0}^{n} 
\frac {z_{1}^{(n-r)}~~z_{2}^{r}}{\sqrt{(n-r)!~r!}} ~~
\vert n-r, r>  
\label{cs2}
\eea

\noindent In (\ref{cs2}), the delta function is the spin projection 
operator and it projects out all states with fixed spin value. 
 $(z_1, z_2)$ is a doublet of complex numbers and satisfy 
the constraint $|z_1|^2 + |z_2|^2 =1$. Thus the coherent states in (\ref 
{cs2}) are defined over the hyper-sphere $S^{3}$. By construction, 
the  coherent states in (\ref{cs2}) satisfy: 
\bea 
{\cal C} \vert \vec{z}>_{n} = n ~ \vert \vec{z}>_{n}. 
\label {evc} 
\eea
The equivalence of 
(\ref{st}) and (\ref{cs2}) is trivially shown by considering the coherent 
states in the simplest $j = 1/2$ or equivalently n=1 representation and we 
get:   

\bea 
z_1 \equiv e^{i {\frac{\psi}{2}}} e^{i {\frac{\phi}{2}}} 
cos {\theta \over 2}, ~~ 
z_2 \equiv e^{i {\frac{\psi}{2}}} e^{-i {\frac{\phi}{2}}} 
sin {\theta \over 2}. 
\label{eq}
\eea 

\noindent Substituting the values (\ref{eq}), the Schwinger construction 
(\ref{cs2}) reduces to the standard construction (\ref{st}) given by 
Perelomov \cite{perelomov}. The resolution of identity follows because 
under the SU(2) group transformation operator $\hat{U} \equiv exp i 
\sum_{a=1}^{3} \alpha^{a}J^{a}$,  $(z_1, z_2)$ rotates as a doublet, i.e,: 
\bea 
z_i \rightarrow \big(exp  {i \over 2} \alpha^{a} \sigma^{a} \big)_{ij} z_j. 
\label{tr}
\eea

\noindent Therefore, the operator, 

\bea 
{\cal I}_n \equiv \int_{SU(2)}~d\mu(\vec{z}) \vert 
\vec{z}>_{n} {}_{n}< \vec{z} \vert, 
\label{roi1} 
\eea
with $\int_{SU(2)}~d\mu(\vec{z}) \equiv \int d^{2}\vec{z}~ 
\delta(|\vec{z}|^2-1)$, remains invariant under SU(2) and therefore, 
by Shur's lemma,  
must be proportional to the identity operator. The above simple procedure 
using Schwinger bosons to construct 
SU(2) coherent states with fixed value of the Casimir was generalized to 
SU(N) in \cite{mm3, mmn}. At this stage, we also have the possibility of 
defining SU(2) coherent states which are eigenstates of the third 
component of the angular momentum \cite{bdr} instead of the spin operator. 
We define the SU(2) charge operator ${\cal Q}$ 
to be twice of the third component of the angular momentum, i.e, 

\bea 
{\cal Q}  = a_{1}^{\dagger}  a_{1} - a_{2}^{\dagger}  a_{2}.  
\label{ch}
\eea 
\noindent We further note that: 
\bea 
\big[{\cal Q}, a_{1} a_{2}\big] = 0
\label{csco2}
\eea
\noindent Thus the complete set of commuting operators for SU(2) is 
$\big({\cal Q}, a_{1}a_{2}\big)$. The SU(2) fixed charge (q) coherent 
states are obtained from 
(\ref{cs2}) simply by replacing the spin projection operator  
by fixed charge projection operator, i.e,: 
\bea
\vert z_1, z_2>_{q} & ~=~ & \delta_{[{a_1}^{\dagger}{a_1}
- {a_2}^{\dagger} a_2,q]}
~ \exp \Big( \vec{z} \cdot \vec{a}^\dagger \Big) ~\vert 0,0> \nonumber \\
& = & \sum_{n=0}^{\infty} 
\frac {z_{1}^{(q+n)}~~z_{2}^{n}}{\sqrt{(q+n)!~n!}} ~~
\vert q+n, n>  \nonumber \\ 
&=& \frac{{z_1}^{q}}{\sqrt{q!}}~ exp~\Big( z_1 z_2~ 
\frac{1}{\hat{N}_1}~{a_1}^{\dagger}{a_2}^{\dagger}\Big) \vert q, 0> 
\label{cs3}
\eea
\noindent Fixed charge coherent states in (\ref{cs3}) satisfy: 
\bea 
{\cal Q} \vert z_1, z_2>_{q} & = & q ~ \vert z_1, z_2>_{q}  
\label {evq} 
\eea
\noindent Thus, the equation (\ref{evq}) for fixed charge replaces 
the equation (\ref{evc}) for the fixed spin coherent states. 
The two types of coherent states defined in (\ref{cs2}) and (\ref{cs3}) 
are very different in their structures.  The projection operator 
(\ref{ch}), unlike Casimir projection operator 
(\ref{cas}), does not commute with the SU(2) generators. Therefore, 
$\vert z_1, z_2>_{j}$  are the vectors in the $j^{th}$ irreducible 
representation of SU(2) while $\vert z_1, z_2>_{q}$ are 
the vectors in the infinite dimensional Hilbert space of the two 
harmonic oscillators. Further, $\vert z_1, z_2>_{q}$ are the eigenstates 
of the destruction operators: 
\bea 
a_{1} a_{2} \vert z_1, z_2>_{q} & = & z_{1} z_{2} \vert z_1, z_2>_{q}
\label{evq2}
\eea
and are defined on $C^{2}$ to satisfy resolution of identity. In this 
sense, (\ref{cs3})  are closer to (\ref{cs1}). To prove the resolution 
of identity, we start with the trivial identity 
operator in the direct product Hilbert space of two Heisenberg-Weyl coherent 
states (\ref{cs1}) characterized by $z_1$ and $z_2$ , i.e,: 
\bea 
{\cal I} \equiv 
\sum_{n_{1},n_{2}=0}^{\infty} 
\sum_{m_{1},m_{2}=0}^{\infty} 
\int d\mu{(z_{1})} \int d\mu{(z_{2})} 
\frac {z_{1}^{n_{1}}~~z_{2}^{n_{2}}} {\sqrt{n_{1}!~~n_{2}!}} ~~
\frac {\bar{z}_{1}^{m_{1}}~~\bar{z}_{2}^{m_{2}}} {\sqrt{m_{1}!~~m_{2}!}} ~~
\vert n_{1}> \vert n_{2}> <m_{1} \vert <m_{2}\vert  
\label{roidp} 
\eea 
In (\ref{roidp}), we insert the following two trivial identities: 
\bea
\sum_{q=0}^{\infty} \delta(n_1-n_2-q) = 1 ~~~~
\sum_{q^{\prime}=0}^{\infty} \delta(m_1-m_2-q^{\prime}) = 1   
\label{ti} 
\eea 
\noindent and rearrange the summations over $(n_{1},n_{2})$ and
$(m_{1},m_{2})$ in terms of q and $q^{\prime}$. The two angular 
integrations over the phases of $z_{1}$ and $z_{2}$  lead to: 
\bea 
{\cal I} \equiv \sum_{q=0}^{\infty} \int d\mu{(z_{1})} 
\int d\mu{(z_{2})} \vert z_1, z_2>_{q} {}_{q} < z_1, z_2\vert 
\label{roifc} 
\eea 

\subsection{Nonlinear Formulation} 

Motivated by the non-linear formulation of the Heisenberg Weyl coherent 
state in (\ref{nlhw}), we can define the non-linear SU(2) 
coherent states as: 
\bea
f(\hat{N}_1,\hat{N}_2)a_1 a_2 \vert z_1, z_2; f>_{q} \equiv  {z_1}{z_2} 
\vert z_1, z_2; f>_{q}
\label{nl2} 
\eea 
We expand $\vert z_1, z_2; f>_{q}$: 
\bea 
\vert z_1, z_2; f>_{q}  \equiv \sum_{n=0}^{\infty} C_{n}(z_1,z_2;f) 
~\vert q+n, n>  
\label{nlsu2} 
\eea
The equation (\ref{nl2}) implies:
\bea 
C_{n}(z_1,z_2;f) = \sqrt{\frac{q~!}{(n+q)!~ n!}}~  
~\prod_{k=1}^{n} ~\frac{z_1~z_2}{f(n+q-k,n-k)}~ C_{0}(z_1,z_2;f)
\label{cnlsu2}
\eea
We would like to put (\ref{nlsu2}) in the exponential form analogous to 
(\ref{nlhwe2}). For this we substitute (\ref{cnlsu2}) in (\ref{nlsu2}) 
and after rearranging terms we get:
\bea
\vert z_1, z_2; f>_{q} & = & \sum_{n=0}^{\infty} \frac{q~!}{(n+q)!~n!}
~\Big(\prod_{i=1}^{2}z_i~a^{\dagger}_i\Big)^{n}~\prod_{l=1}^{n}
\frac{1}{f(\hat{N}_1+n-l,\hat{N}_2+n-l)} ~\vert q, 0> \nonumber \\
&\equiv& \sum_{n=0}^{\infty} \frac{1}{n!}~ 
\Big(\prod_{i=1}^{2}z_i~a^{\dagger}_i\Big)^{n} ~\Big[\prod_{k=1}^{n} 
\hat{H}_{f}(\hat{N}_{1}+k,\hat{N}_2+k)\Big] ~\vert q, 0>  
\label{abc}
\eea
where, 
\bea 
\hat{H}_{f}(\hat{N}_{1}+k, \hat{N}_{2}+k) \equiv \frac{1}{(\hat{N}_1+k)} 
\frac{1}{f(\hat{N}_1-1+k, \hat{N}_{2}-1+k)}.    
\label{hh2}
\eea
Now using the operator identity: 
\bea 
\Big(a_1^{\dagger} a_2^{\dagger}\Big)^{n} \prod_{k=1}^{n} 
\hat{H}_f(\hat{N}_1+k, \hat{N}_{2}+k) = \Big[\hat{H}_f(\hat{N}_1, \hat{N}_2) 
a_1^{\dagger} a_2^{\dagger}\Big]^{n},  
\label{id4} 
\eea
we finally get: 
\bea 
|z_1, z_2; f>_{q} = C_{0}~ exp~\Big[z_1z_2~\hat{H}_f(\hat{N}_1,
\hat{N}_2)~a_1^{\dagger}a_2^{\dagger}\Big] 
~\vert q, 0> 
\label{exp45} 
\eea 
Using the condition: $|z_1, z_2; f=1>_q = |z_1, z_2>_q$, we get $C_{0} 
= {{z_1^q} \over {\sqrt{q!}}}$. We note that the structure of 
SU(2) coherent states $|z_1, z_2; f>_{q}$ in (\ref{exp45}) is same as 
that of the Heisenberg Weyl coherent state $|z, f \rangle_{\infty}$ 
in (\ref{nlhwe2}).

\noindent We thus see in the above sections that the Schwinger boson algebra 
associated with SU(2) group is extremely useful and convenient to 
define SU(2) coherent states with fixed spin (\ref{cs2})  
as with fixed charges (\ref{cs3}).  Further, one can trivially generalize 
the above construction to define fixed charge 
nonlinear coherent states. At this stage, it should be emphasized that 
for fixed charge linear and non-linear coherent states, the Schwinger 
boson construction presented here is most natural. This is because it involves 
the operators $a_1, a_2$, e.g the equations (\ref{evq2}) and (\ref{nl2}), 
which are  like `square roots' of the angular momentum operators. Further,  
it is not obvious how a simple modification of the standard construction 
(\ref{st}) can lead to fixed charge coherent states. 
In the next two section we  will generalize these ideas to SU(N) group
and construct SU(N) coherent states with fixed (N - 1) charges. 

\section{\bf Coherent States with SU(3) Charges}

\noindent The group SU(3) is of rank two and therefore it has two 
fundamental representations transforming like triplet and anti-triplet. 
We, therefore,  define two triplets of harmonic oscillator creation and
annihilation operators $(a_i,b_i)$, i=1,2,3, satisfying \cite{georgi}:
\bea 
\big[a_i,a^{\dagger}_j \big] = & \delta_{ij} ~, ~~& \quad
\big[b_i,b^{\dagger}_j \big] = \delta_{ij} ~, \nonumber \\ 
\big[a_i,b_j\big] = & 0 ~,~~ & \quad \big[a_i,b^{\dagger}_j \big] = 0 ~.
\eea
We will  denote these two triplets by $(\vec{a},\vec{b})$ 
and the two number operators by $\hat{N}_{i}$ and $\hat{M}_{i}$ 
respectively; $\hat{N}_i \equiv {a_{i}}^{\dagger}a_{i}, \hat{M}_i 
\equiv {b_{i}}^{\dagger}b_{i}$. 
Similarly, the vacuum state can be labelled by $|\vec{0}_a,\vec{0}_b>$. 
Henceforth, we will ignore the subscripts $a, b$ and will denote the 
vacuum state by $|\vec{0},\vec{0}>$ or more explicitly by 
$\vert {}^{0,~0,~0}_{0,~0,~0}\rangle$. 
Let $\frac{\lambda^a}{2}$, $a=1,2,...,8$ be the generators of $SU(3)$ in the 
fundamental representation; they satisfy the $SU(3)$ Lie algebra 
$[\frac{\lambda^a}{2}, \frac{\lambda^b}{2}] = 
i f^{abc} \frac{\lambda^c}{2}$. Let us define the following operators
\beq
Q^a = \frac{1}{2} \Big(a^\dagger \lambda^a a - b^\dagger \lambda^{*a} b
\Big) ~,
\label{suc} 
\eeq
where $a^\dagger \lambda^a a \equiv a^{\dagger}_i \lambda^a_{ij} 
a_j$, and $b^\dagger \lambda^{*a} b \equiv b^{\dagger}_i 
\lambda^{*a}_{ij} b_j$. To be explicit,
\bea
Q^3 ~&=&~ \frac{1}{2} ~( a_1^\dagger a_1 ~-~ a_2^\dagger a_2 ~-~ b_1^\dagger 
b_1 ~+~ b_2^\dagger b_2 ) ~, \nonumber \\
Q^8 ~&=&~ \frac{1}{2 {\sqrt 3}} ~( a_1^\dagger a_1 ~+~ a_2^\dagger a_2 ~-~ 2 
a_3^\dagger a_3 ~-~ b_1^\dagger b_1 ~-~ b_2^\dagger b_2 ~+~ 2 b_3^\dagger 
b_3 ) ~, \nonumber \\
Q^1 ~&=&~ \frac{1}{2} ~( a_1^\dagger a_2 ~+~ a_2^\dagger a_1 ~-~ b_1^\dagger 
b_2 ~-~ b_2^\dagger b_1 ) ~, \nonumber \\
Q^2 ~&=&~ -\frac{i}{2} ~( a_1^\dagger a_2 ~-~ a_2^\dagger a_1 ~+~ b_1^\dagger 
b_2 ~-~ b_2^\dagger b_1 ) ~, \nonumber \\
Q^4 ~&=&~ \frac{1}{2} ~( a_1^\dagger a_3 ~+~ a_3^\dagger a_1 ~-~ b_1^\dagger 
b_3 ~-~ b_3^\dagger b_1 ) ~, \nonumber \\
Q^5 ~&=&~ -\frac{i}{2} ~( a_1^\dagger a_3 ~-~ a_3^\dagger a_1 ~+~ b_1^\dagger 
b_3 ~-~ b_3^\dagger b_1 ) ~, \nonumber \\
Q^6 ~&=&~ \frac{1}{2} ~( a_2^\dagger a_3 ~+~ a_3^\dagger a_2 ~-~ b_2^\dagger 
b_3 ~-~ b_3^\dagger b_2 ) ~, \nonumber \\
Q^7 ~&=&~ -\frac{i}{2} ~( a_2^\dagger a_3 ~-~ a_3^\dagger a_2 ~+~ b_2^\dagger 
b_3 ~-~ b_3^\dagger b_2 ) ~.
\label{su3}
\eea
It can be checked that these operators satisfy the $SU(3)$ algebra amongst 
themselves, i.e., $[Q^a,Q^b] = i f^{abc} Q^c$. 
The two Casimirs of SU(3) are the two types of total number 
operators, i.e,: 
\bea 
{\cal C}_{1} & = &  \hat{N}_1 +  \hat{N}_2 +  \hat{N}_3, \nonumber \\ 
{\cal C}_{2} & = &  \hat{M}_1 +  \hat{M}_2 +  \hat{M}_3. 
\label{su3casimir}
\eea
This can be seen simply by looking at the SU(3) generators in (\ref{su3}). 
They always involve one creation and one destruction operators of each 
type and hence both the operators in (\ref{su3casimir}) commute with 
all the SU(3) generators.  The eigenvalues of ${\cal C}_1$ and ${\cal C}_2$ 
will be denoted by n and m respectively. Further, 
\bea
\big[Q^a, a^{\dagger}_i \big] ~&=&~ \lambda^a_{ji} a^{\dagger}_j ~, \quad 
\quad ~~ \big[Q^a, b^{\dagger}_i\big] ~ = - \lambda^{*a}_{ji} b^{\dagger}_j ~,
\nonumber \\
\hspace{1.0cm} 
\big[Q^a, a^\dagger \cdot a \big] ~&=&~~~ 0 ~~~~, ~~ \quad \quad \big[Q^a, 
b^\dagger \cdot b\big] ~~~ = 0 ~~~
\label{cass} 
\eea
From Eqs. (\ref{cass}), it is clear that the three states 
$a^{\dagger}_i |
\vec{0},\vec{0}>$ with ($n=1,m=0$) and $b^{\dagger}_i |\vec{0},\vec{0}>$ with 
($n=0, m=1$) transform respectively as the fundamental representation $3$ 
and its conjugate $\bar{3}$ representation.  
By taking the direct product of 
$N$ ${\vec{a}}^\dagger$'s and M ${\vec{b}}^\dagger$'s and subtracting 
the traces we can form any of the higher irreducible representations of SU(3) 
\cite{mm3}. We can now define the SU(3) coherent states with fixed values 
of Casimirs (or generalized coherent states) as: 
\bea 
|\vec{z},\vec{w}>_{(n,m)} ~\equiv ~ \delta_{({\cal C}_1,n)}
\delta_{({\cal C}_2,m)} \exp ~(\vec{z} \cdot 
\vec{a}^\dagger ~+~ \vec{w} \cdot \vec{b}^\dagger ) 
|\vec{0},\vec{0}> 
\label{csc} 
\eea
The coherent states in (\ref{csc}) are straight forward generalization 
of the corresponding SU(2) coherent states in (\ref{cs2}). The two delta 
function constraints mean: 
\bea 
{\cal C}_{1}~|\vec{z},\vec{w}>_{(n,m)} & = & n~|\vec{z},\vec{w}>_{(n,m)} 
\nonumber \\
{\cal C}_{2}~|\vec{z},\vec{w}>_{(n,m)} & = & m~|\vec{z},\vec{w}>_{(n,m)} 
\label{cascon3}
\eea 
The  complex vectors $(\vec{z},\vec{w})$ in (\ref{csc}) are the sets of 
two orthonormal triplets \cite{mm3}  describing the SU(3) manifold. 
They satisfy: 
\bea 
|\vec{z}|^2 = |\vec{w}|^2 = 1, ~~~ \vec{z}.\vec{w} = 0. 
\label{3man}
\eea
The resolution of identity follows from the fact that under SU(3) 
transformations $\vec{z}$ and $\vec{w}$ transform like triplet and 
anti-triplet respectively. Therefore, the operator: 
\bea 
{\cal I}_{(n,m)} \equiv \int_{SU(3)} d\mu(\vec{z},\vec{w}) 
~~\vert \vec{z}, \vec{w}>_{(n,m)}~ 
{}_{(n,m)}< \vec{z}, \vec{w} \vert, 
\label{roi3} 
\eea
with $\int_{SU(3)} d\mu(\vec{z},\vec{w}) \equiv 
\int d^{3}\vec{z}~d^{3}\vec{w} \delta(|\vec{z}|^2-1)\delta(|\vec{w}|^2-1)
\delta(\vec{z}.\vec{w})$, is an SU(3) invariant. Thus ${\cal I}_{(n,m)}$
in (\ref{roi3}) commutes with all the SU(3) generators in (\ref{su3}) and 
hence it is proportional to the identity operator for all possible values of 
(n,m).  On the other hand, the SU(3) Lie algebra involves two diagonal 
generators and therefore we can define two mutually commuting charge operators: 
\bea 
{\cal Q}_1 \equiv 2 ~ Q^{3}, ~~~ {\cal Q}_2 \equiv 2 \sqrt{3} Q^{8}  
\label{co} 
\eea

%We now define the charge and hypercharge operators as:
%\bea
%{\cal Q} & \equiv &  Q^{3} + \frac{1}{\sqrt{3}} Q^{8} = \frac{1}{3}\big(
%2a_{1}^{\dagger}a_{1}-a_{2}^{\dagger}a_{2} - a_{3}^{\dagger}a_{3} -
%2b_{1}^{\dagger}b_{1}+b_{2}^{\dagger}b_{2} + b_{3}^{\dagger}b_{3} \big)
%  \nonumber \\
%{\cal Y} & = & \frac{2}{\sqrt{3}} Q^{8} = \frac{1}{3}\big(
%a_{1}^{\dagger}a_{1}+a_{2}^{\dagger}a_{2} - 2a_{3}^{\dagger}a_{3} -
%b_{1}^{\dagger}b_{1}-b_{2}^{\dagger}b_{2} + 2b_{3}^{\dagger}b_{3} \big) 
%\label{cy}
%\eea 
% 
%We now define the SU(3) manifold can be defined in terms of two 
%complex triplets $(\vec{z},\vec{w})$ satisfying \cite{md}:
%\bea
%|\vec{z}|^{2} = |\vec{w}|^{2} =1~, \vec{z}.\vec{w} = 1. 
%\label{cons} 
%\eea 
\noindent It is easy to see that the complete set of commuting operators 
(CSCO) is given by  the set\footnote{Note that the formulation in 
\cite{chin} corresponds to ignoring $b_1, b_2$ and $b_3$ or equivalently 
 putting $M_1=M_2=M_3 =0$.}:  
$\big[{\cal Q}_1, {\cal Q}_2, a_{1}a_{2}a_{3}, b_{1}b_{2}b_{3}, a_{1}b_{1}, 
a_{2}b_{2}, a_{3}b_{3}\big]$.  We now define the coherent states with fixed 
SU(3) charges as:  
\beq
|\vec{z},\vec{w}>_{(q_1,q_2)} ~\equiv ~ \delta_{({\cal Q}_1,q_{1})}
\delta_{({\cal Q}_2,q_{2})} \exp ~(\vec{z} \cdot 
\vec{a}^\dagger ~+~ \vec{w} \cdot \vec{b}^\dagger ) 
|\vec{0},\vec{0}> 
\label{csf} 
\eeq
In (\ref{csf}), the two $\delta$ function constraints imply that the 
states on the left satisfy 
\bea 
{\cal Q}_1 \vert \vec{z},\vec{w} \rangle_{(q_{1},q_{2})} 
& = & q_1~ \vert \vec{z},\vec{w} \rangle_{(q_1,q_2)} \nonumber 
\\
{\cal Q}_2 \vert \vec{z},\vec{w} \rangle_{(q_{1},q_{2})} 
& = & q_2~ \vert \vec{z},\vec{w} \rangle_{(q_1,q_2)}  
\label{eve3} 
\eea 
\noindent The equations (\ref{csf}) and (\ref{eve3}) are the SU(3) 
analogs of the similar equations (\ref{cs3}) and (\ref{evq}) in 
the case of SU(2) respectively.  
In terms of the occupation number 
states characterized by $({{n}_1, {n}_2, {n}_3})$ and $(m_1, m_2, m_3)$, 
the eigenvalue equations (\ref{eve3}), 
\bea 
{\cal Q}_a\vert {}^{{n}_1,~{n}_2,~{n}_3}_{m_1, m_2, m_3}\rangle
= q_{a} ~\vert {}^{{n}_1,~{n}_2,~{n}_3}_{m_1, m_2, m_3}\rangle, ~~ a=1, 2 
\label{neve} 
\eea
\noindent can be easily solved for the first two occupation numbers 
$(n_1, n_2)$ in terms of the two SU(3) charges $(q_1, q_{2})$ and 
rest of the occupation numbers.  The solutions are:

\bea
\label{sol3}
n_{2} & = & n_3 + \frac{1}{2} (q_2 - q_1) + 2(m_2-m_3) \\
n_{1} &  = & n_3 + q_1 + \frac{1}{2} (q_2 - q_1) + (m_1-m_2)+2(m_2 - m_3) 
\nonumber 
\eea  
\noindent By defining $q_{0} \equiv 0$, the solutions in (\ref{sol3}) 
can be written in a compact form: 

\bea 
n_{i} = n_3 + l_i 
+ \sum_{a=i}^{2}~ a ~(m_a-m_{a+1}), ~~~ i=1, 2.  
\label{cf}
\eea 
In (\ref{cf}), $l_i \equiv \sum_{a=i}^{2} \Big(\frac{1}{a}(q_a-q_{a-1})$. 
The above solution for SU(3) group is written in a form which can be 
generalized to SU(N) group (see section 4 ). 
The equations (\ref{csf}) and (\ref{eve3}) imply: 
\bea
|\vec{z},\vec{w}>_{(q_1,q_2)} = 
{\sum_{{n}_3 = 0}^\infty}~~ {\sum_{m_{1},m_{2},m_{3} = 0}^{\infty}}
B^{{n}_3}_{m_1,m_2,m_3}(\vec{z},\vec{w}) \vert {}^{{n}_1,~{n}_2,~{n}_3}
_{m_1, m_2, m_3}\rangle   
\label{exp1} 
\eea 
\noindent where, 
\bea
B^{{n}_3}_{m_1,m_2,m_3}(\vec{z},\vec{w}) \equiv \frac{z_{1}^{{n}_1} 
z_{2}^{{n}_2} z_{3}^{{n}_3} w_{1}^{m_{1}} w_{2}^{m_{2}} w_{3}^{m_{3}}}{
\Big[{n}_1! {n}_2! {n}_3! m_1! m_2! m_3!  \Big]^{\frac{1}{2}}} 
\label{exp2} 
\eea 

\noindent In (\ref{exp2}), the occupation numbers $n_{1}$ and $n_2$ 
are given in terms of the two charges $q_1$, $q_2$ and the rest of 
the occupation numbers by (\ref{sol3}). We further note 
that the coherent states defined in (\ref{exp1}) are infact eigenstates 
of CSCO, i.e,: 
\bea
{{\cal Q}_a} \vert\vec{z},\vec{w} \rangle_{(q_1,q_2)} & = & q_{a} 
\vert\vec{z},\vec{w} \rangle_{(q_1,q_2)}, ~~~~~~~~~~ a=1,2 \nonumber \\
\label{eve2} 
a_{1}a_{2}a_{3} \vert\vec{z},\vec{w} \rangle_{(q_1,q_2)} 
& = & z_{1}z_{2}z_{3} 
\vert \vec{z},\vec{w} \rangle_{(q_1,q_2)},  \\
b_{1}b_{2}b_{3} \vert \vec{z}, \vec{w} \rangle_{(q_1,q_2)} 
&  = & w_{1}w_{2}w_{3}  \vert \vec{z},\vec{w} \rangle_{(q_1,q_2)} \nonumber  
 \\ 
a_{i}b_{i} \vert \vec{z},\vec{w} \rangle_{(q_1,q_2)}  & = &  z_{i}w_{i} 
\vert \vec{z},\vec{w} \rangle_{(q_1,q_2)},~~~~~~~~ for~ each ~ i~ (=1, 2, 3)  
\nonumber 
\eea 
\noindent The above eigenvalue equations are the SU(3) analogs of 
the corresponding SU(2) equations in (\ref{evq}). 
The  coherent states in (\ref{exp1}) can also 
be written as\footnote{ The proof is given in the next section.}: 
\bea
|\vec{z},\vec{w}>_{(q_1,q_2)} = 
exp~\Big[\frac{z_1z_2z_3}{\hat{N}_1\hat{N}_2} 
~a_1^{\dagger}a_2^{\dagger}a_3^{\dagger} + \frac{w_1w_2w_3}{M_1M_2}
b_1^{\dagger}b_2^{\dagger}b_3^{\dagger}\Big] 
\sum_{\nu_1,\nu_2= 0}^{\infty} C^{0}_{\nu_1,\nu_2,0}
\vert {}^{l_1+\nu_1 ~~ l_2
+\nu_2~~~0} _{~~ \nu_1~~~~~\nu_2~~~~\phantom{.} 0}~\rangle   
\label{exp44}
\eea 
\noindent In (\ref{exp44}): 
\bea 
C^{0}_{\nu_1,\nu_2,0} \equiv \frac{z_1^{l_1+\nu_1}
z_2^{l_2+\nu_2}
w_1^{\nu_1} w_2^{\nu_2}}{\sqrt{(l_1+\nu_1)!
(l_2+\nu_2)!\nu_1!\nu_2!}} 
\label{def} 
\eea
Further, the two summations in (\ref{exp44}) can be performed to get the 
final compact exponential form (see section 3.1 ): 
\bea 
|\vec{z},\vec{w}>_{(q_1,q_2)}   =  \frac{{z_1}^{l_1}}{\sqrt{l_1!}} 
\frac{{z_2}^{l_2}}{\sqrt{l_2!}} 
exp~\Big[{z_1z_2z_3}\frac{1}{\hat{N}_1 \hat{N}_2}
a_1^{\dagger}a_2^{\dagger}a_3^{\dagger} +   
{w_1w_2w_3}\frac{1}{\hat{M}_1\hat{M}_2}
b_1^{\dagger}b_2^{\dagger}b_3^{\dagger}\Big] 
\nonumber \\
exp~\Big[z_1w_1\frac{1}{\hat{N}_1}a_1^{\dagger}b_1^{\dagger} 
+ z_2w_2\frac{1}{\hat{N}_2}a_2^{\dagger}b_2^{\dagger}\Big]  
\vert {}^{l_1 ~~~ l_2~~~0} _{0~~~~0~~~ 0}~\rangle   
\label{exp88} 
\eea
\noindent The proof for the resolution of identity for the fixed charge 
SU(3) coherent states (\ref{exp44}) is similar to that for the case of 
SU(2).  We start with a direct product of 6 Heisenberg-Weyl coherent states: 
\bea 
\label{csnn} 
|\vec{z},\vec{w}>_{_{HW}} ~& \equiv& ~ \exp ~(\vec{z} \cdot 
\vec{a}^\dagger ~+~ \vec{w} \cdot \vec{b}^\dagger ) 
|\vec{0},\vec{0}>  \\  
&=& {\sum_{n_1,n_2,n_3 = 0}^\infty}~~ {\sum_{m_{1},m_{2},m_{3} = 0}^{\infty}}
A^{n_1,~n_2,~n_3}_{m_1,m_2,m_3}(\vec{z},\vec{w})  
\vert {}^{n_1,~n_2,~n_3}_{m_1, m_2, m_3}\rangle   
\nonumber
\eea 
\noindent In (\ref{csnn}), 
\bea
A^{n_1,~n_2,~n_3}_{m_1,m_2,m_3}(\vec{z},\vec{w})  \equiv  
\frac{z_{1}^{n_{1}} z_{2}^{n_2} z_{3}^{n_3} 
w_{1}^{m_{1}} w_{2}^{m_{2}} w_{3}^{m_{3}}}{\Big[n_1! n_2! n_3! m_1! m_2! m_3!
\Big]^{\frac{1}{2}}}  
\label{cc}
\eea
\noindent The coherent states defined by (\ref{csnn}) trivially satisfy the 
resolution of identity: 
\bea
\int d\mu(\vec{z},\vec{w}) 
|\vec{z},\vec{w}\rangle_{_{_{HW}}} {}_{_{_{HW}}} 
\langle\vec{z},\vec{w} \vert \equiv  {\cal{I}} 
\label{roi} 
\eea 
\noindent where, $d\mu(\vec{z},\vec{w}) \equiv 
\Big[\prod_{i=1}^{3}~ d^{2}z_{i}~d^{2}w_{i}\Big] ~ e^{-(|\vec{z}|^{2} 
+ |\vec{w}|^{2})}$. 
%The equations (\ref{suc}) and (\ref{cs3}) show that under  a SU(3) 
%transformation, the vectors $\vec{z}$ and $\vec{w}$ transform as 
%triplet and anti-triplet respectively, i.e, 
%\bea 
%\vec{z} & \longrightarrow & exp i (\theta^{a} \lambda^{a}) \vec{z} \nonumber \\
%\vec{w} & \longrightarrow &  exp - i (\theta^{a} {\lambda^{*}}^{a}) \vec{w}
%\label{trans} 
%\eea 
We now expand ${\cal{I}}$ in (\ref{roi})  as:  
\bea 
\int d\mu(\vec{z},\vec{w}) {\sum_{\vec{n} = 0}^\infty}~ {\sum_{\vec{m} = 
0}^{\infty}}~~
{\sum_{\vec{l} = 0}^\infty}~ {\sum_{\vec{p} = 0}^{\infty}}
A^{n_1,~n_2,~n_3}_{m_1,m_2,m_3}(\vec{z},\vec{w}) {A^{*}}^{l_1,~l_2~,~l_3}_
{p_1,~p_2,~p_3}(\vec{z},\vec{w})~~ 
\vert {}^{n_1,~n_2,~n_3}_{m_1, m_2, m_3}\rangle   
\langle {}^{l_1,~l_2,~l_3}_{p_1,~p_2,~p_3}\vert 
\nonumber 
\eea   
\noindent and rearrange the summations by inserting the following 
identities: 
\bea
\sum_{q_1} \delta\big((n_1-n_2)-(m_1-m_2)-q_1\big) &=& 1
\nonumber  \\ 
\sum_{q_2} \delta\big((n_1+n_2-2n_3)-(m_1+m_2-2m_3)-q_2\big) & = & 1,  
\nonumber  \\ 
\sum_{{q_1}^{\prime}} \delta\big((l_1-l_2)-(p_1-p_2)-{q_1}^{\prime}
\big) &=& 1
\nonumber \\
\sum_{{q_2}^{\prime}} \delta\big((l_1+l_2-2l_3)-(p_1+p_2-2p_3)-
{q_2}^{\prime} \big) &=& 
1. 
\label{ids}
\eea
\noindent The identities in (\ref{ids}) are SU(3) analogs of 
(\ref{ti}) and  like in 
the case of SU(2), the angular integrations lead to: 
\bea 
{\cal{I}} & = & \sum_{q_1,q_2} \sum_{{n}_3=0}^{\infty}\sum_{\vec{m}=0}^{\infty} 
\int d\mu({r},{R}) \prod_{i=1}^3\frac{{(r_{i}^2)}^{{n}_i}{(R_{i}^2)}^{m_i}} 
%(r_1^2)^{n_1} (r_2^2)^{n_2} (r_3^2)^{n_3} (R_1^2)^{m_1} 
%(R_2^2)^{m_2} (R_3^2)^{m_3}}
{{n}_i! m_i!} 
\vert{}^{{n}_1,~{n}_2,~{n}_3}_{m_1,m_2,m_3}\rangle 
\langle{}^{{n}_1,~{n}_2,~{{n}_3}}_{m_1,m_2,m_3}\vert \nonumber \\ 
&=& \sum_{q_1,q_2} \int d\mu(\vec{z},\vec{w}) \vert \vec{z},\vec{w} 
\rangle{}_{(q_1,q_2)}~ 
{}_{(q_1,q_2)} \langle \vec{z},
\vec{w} \vert
\label{ooo} 
\eea 
\noindent In (\ref{ooo}), $({n}_1, {n}_2)$  are given in terms 
of charges $(q_1, q_2)$ by  equations (\ref{sol3}). 

\subsection{Nonlinear Formulation} 

\noindent We define the non-linear SU(3) charge  coherent states 
as the common eigenvectors of the following operators: 
\bea 
f(\hat{N}_1,\hat{N}_2,\hat{N}_3) a_1a_2a_3 \vert\vec{z},\vec{w},f,g \rangle_{q_1,q_2} 
& = & z_1 z_2 z_3 \vert\vec{z},\vec{w},f,g \rangle_{q_,q_2} \nonumber \\
g(\hat{M}_1,\hat{M}_2,\hat{M}_3) b_1b_2b_3 \vert\vec{z},\vec{w},f,g \rangle_{q_1,q_2}  
& = & w_1 w_2 w_3 \vert\vec{z},\vec{w},f,g \rangle_{q_1,q_2} 
\label{nl}
\eea 

\noindent The number operators in (\ref{nl}) $(\vec{\hat{N}},\vec{\hat{M}})$ have been 
defined in the previous section.  We now define 
\bea
|\vec{z},\vec{w},f,g>_{(q_1,q_2)} = 
{\sum_{{n}_3 = 0}^\infty}~~ {\sum_{m_{1},m_{2},m_{3} = 0}^{\infty}}
C^{{n}_3}_{m_1,m_2,m_3}(\vec{z},\vec{w},f,g) \vert {}^{{n}_1, {n}_2, {n}_3}
_{m_1, m_2, m_3}\rangle   
\label{exp3} 
\eea 
In (\ref{exp3}) $(n_1,n_2)$ are defined in (\ref{sol3}). The equation 
(\ref{nl}) implies: 
\bea 
C^{n_3}_{m_1,m_2,m_3}(\vec{z},\vec{w},f,g)  =  \frac{z_1 z_2 z_3}
{\sqrt{n_1 n_2 n_3} f(n_1-1,n_2-1,n_3-1)} C^{n_3-1}_{m_1,m_2,m_3}
(\vec{z},\vec{w},f,g) \nonumber \\   
 =  \frac{w_1 w_2 w_3}{\sqrt{m_1 m_2 m_3} 
g(m_1-1,m_2-1,m_3-1)} C^{n_3}_{m_1-1,m_2-1,m_3-1}(\vec{z},\vec{w},f,g)   
\label{exp4} 
\eea
\noindent Note that: 
\bea 
C^{n_3}_{m_1,m_2,m_3}(\vec{z},\vec{w},f=1,g=1) = B^{n_3}_{m_1,m_2,m_3}
(\vec{z},\vec{w}) 
\label{smr} 
\eea 

\noindent where $B^{n_3}_{m_1,m_2,m_3}(\vec{z},\vec{w})$ are defined in 
equation (\ref{exp2}). Iterating equation (\ref{exp4}), we find: 
\bea
\label{exp5} 
&& C^{n_3}_{m_1,m_2,m_3}(\vec{z},\vec{w},f,g)  = 
\sqrt{\frac{(n_1-n_3)!(n_2-n_3)!(m_1-m_3)!(m_2-m_3)!}
{n_1!n_2!n_3!  m_1!m_2!m_3!}} \\
&& \prod_{k=1}^{n_3} \frac{z_1~z_2~z_3}{f(n_1-k,n_2-k,n_3-k)} 
\prod_{l=1}^{m_3} 
\frac{w_1~w_2~w_3}{g(m_1-l,m_2-l,m_3-l)} 
C^{0}_{m_1-m_3,m_2-m_3,0}(\vec{z},\vec{w},f,g) \nonumber   
\eea 
\noindent The equation (\ref{smr}) implies: 
\bea
 C^{0}_{\nu_1,\nu_2,0}(\vec{z},\vec{w},1,1) =  C^{0}_{\nu_1,\nu_2,0} = 
\frac{z_{1}^{l_1+\nu_1}
z_{2}^{l_2+\nu_2}w_{1}^{\nu_1}w_{2}^{\nu_2}}
{\sqrt{(l_1+\nu_1)!(l_2+\nu_2)!
\nu_1!\nu_2!}}  
\label{bc}
\eea
In (\ref{bc}) $  C^{0}_{\nu_1,\nu_2,0} $ is given by the equation 
(\ref{exp44}). Substituting $C^{n_3}_{m_1,m_2,m_3}$ from (\ref{exp5}) 
into  (\ref{exp3}) and after some algebra, we get: 
\bea
|\vec{z},\vec{w},f,g>_{(q_1,q_2)} = \sum_{\nu_1,\nu_2= 0}^{\infty}
C^{0}_{\nu_1,\nu_2,0} \sum_{m_3 = 0}^{\infty} 
\frac{1}{m_3!}\Big(\prod_{j=1}^{3}w_{j} b_{j}^{\dagger}
\Big)^{m_3} \prod_{l=1}^{m_3} \hat{H}_{g}(M_1+l,M_2+l,M_3+l) \nonumber \\   
{\sum_{{n}_3 = 0}^\infty} \frac{1}{n_3!} 
\big(\prod_{i=1}^{3}z_{i} a_{i}^{\dagger}\big)^{n_3}  
\prod_{k=1}^{n_3}
\hat{H}_{f}(\hat{N}_1+k,\hat{N}_2+k,\hat{N}_3+k) 
\vert {}^{l_1+\nu_1 ~~~ l_2+\nu_2~~~0} _{~~ \nu_1~~~~~~~\nu_2~~~\phantom{i} 0}~\rangle   
\label{exp6} 
\eea
\noindent In (\ref{exp6}), $\nu_1=m_1-m_3, \nu_2=m_2-m_3$ and the operator 
$\hat{H}$ is given by:  
\bea
\hat{H}_{f}(\hat{N}_1,\hat{N}_2,\hat{N}_3) &=& \frac{1}{\hat{N}_1 \hat{N}_2~f(\hat{N}_1-1,\hat{N}_2-1,\hat{N}_3-1)} \nonumber \\
\hat{H}_{g}(\hat{M}_1,\hat{M}_2,\hat{M}_3) &=& \frac{1}{\hat{M}_1 \hat{M}_2~g(\hat{M}_1-1,\hat{M}_2-1,\hat{M}_3-1)} 
\label{id} 
\eea 
\noindent Using the operator identities: 
\bea
\Big(a_1^{\dagger} a_2^{\dagger} a_3^{\dagger}\Big)^{n_3} \prod_{k=1}^{n_3} 
\hat{H}_f(\hat{N}_1+k,\hat{N}_2+k,\hat{N}_3+k) & = & \Big[\hat{H}_f(\hat{N}_1,\hat{N}_2,\hat{N}_3) 
a_1^{\dagger} a_2^{\dagger} a_3^{\dagger}\Big]^{n_3} \nonumber \\  
\Big(b_1^{\dagger} b_2^{\dagger} b_3^{\dagger}\Big)^{m_3} \prod_{l=1}^{m_3} 
\hat{H}_g(\hat{M}_1+l,\hat{M}_2+l,\hat{M}_3+l) & = & \Big[\hat{H}_g(\hat{M}_1,\hat{M}_2,\hat{M}_3) 
b_1^{\dagger} b_2^{\dagger} b_3^{\dagger}\Big]^{m_3}  
\label{id2} 
\eea 
\noindent in equation (\ref{exp6}), we get: 
\bea 
|\vec{z},\vec{w},f,g>_{(q_1,q_2)} ~~  = ~~  
exp~\Big[{z_1z_2z_3}\hat{H}_f a_1^{\dagger}a_2^{\dagger}a_3^{\dagger} +  
{w_1w_2w_3}{\hat{H}_g} b_1^{\dagger}b_2^{\dagger}b_3^{\dagger}\Big] 
\nonumber \\  
 \sum_{\nu_1,\nu_2= 0}^{\infty} C^{0}_{\nu_1,\nu_2,0}~~ 
\vert {}^{l_1+\nu_1 ~~~ l_2+\nu_2~~~0} _{~~ \nu_1~~~~~~~\nu_2~~~\phantom{j} 0}~\rangle   
\label{exp7} 
\eea 
Note that the additional summations over the indeces $\nu_1$ and 
$\nu_2$ in (\ref{exp7}), corresponding to the second fundamental 
representation of SU(3), are abscent for the SU(2) case. These two $\nu_i$
summations can also be exponentiated 
to finally get: 
\bea 
|\vec{z},\vec{w},f,g>_{(q_1,q_2)}   =  C_{0}~  
exp~\Big[{z_1z_2z_3}\hat{H}_f a_1^{\dagger}a_2^{\dagger}a_3^{\dagger} +   
 {w_1w_2w_3}{\hat{H}_g}
b_1^{\dagger}b_2^{\dagger}b_3^{\dagger}\Big] 
\nonumber \\
exp~\Big[z_1w_1\frac{1}{\hat{N}_1}a_1^{\dagger}b_1^{\dagger} 
+ z_2w_2\frac{1}{\hat{N}_2}a_2^{\dagger}b_2^{\dagger}\Big]  
\vert {}^{l_1 ~~~ l_2~~~0} _{0~~~~0~~~ 0}~\rangle   
\label{exp78} 
\eea
Comparing (\ref{exp78}) with (\ref{exp88}) for f=g =1, we get: 
$C_0 = \frac{{z_1}^{l_1}{z_2}^{l_2}}{\sqrt{l_1l_2}}$.   
The coherent states in (\ref{exp78}) are the SU(3) analogs of 
(\ref{exp45}) for SU(2).

\section{\bf Coherent States with SU(N) Charges}

\noindent In this section we will find that the SU(3) formulation  of 
section 3 can be recast into SU(N) co-variant formulation by some 
simple substitutions.  The SU(N) group is of rank (N-1) and therefore it has 
(N-1) fundamental representations \cite{georgi}. For it's Lie 
algebra it is convinient to use the Cartan-Weyl basis and generalize 
the notations of section 3 . 
The (N-1) mutually commuting generators belonging to the Cartan subspace 
will be denoted by $H^{a}$, a=1,2,...,(N-1) and the raising and lowering 
operators by $E^{\alpha}, \alpha = 1,2,...,{(N^{2}-N) \over 2}$. The SU(N) 
Lie algebra in the Cartan Weyl basis is: 
\bea 
\left[H^{a},H^{b}\right]   & = &  0, \nonumber \\
\left[H^{a}, E^{\alpha}\right] &  = &  K^{a}(\alpha) E^{\alpha} \nonumber \\
\left[E^{\alpha},E^{\beta}\right]  & = &
\vec{K}(\alpha).\vec{H}
~~~ if ~~~ \alpha = - \beta \nonumber \\
                                    & =  &   
N^{\alpha,\beta}_{\gamma} E^{\gamma} ~~~ otherwise.
\label{cartan}
\eea
In (\ref{cartan}), $\vec{K}(\alpha)$ are  the $(N - 1)$ dimensional root 
vectors corresponding to the ladder operator $E^{\alpha}$. 
$N_{\alpha,\beta}^{\gamma}$ are constants depending on N and are non-zero 
only if $\vec{K}(\alpha) + \vec{K}(\beta) = \vec{K}(\gamma)$. The algebra 
(\ref{cartan}) has (N-1) fundamental representations which will be 
labeled by the index $F$ (= 1,2,....(N-1)) such that the $F^{th}$
representation is of dimension  ${}^{N} C_{F}$. The generators in 
this representation will be denoted by $(H^{a}(F), E^{\alpha}(F))$. 
To construct the first fundamental representation (F=1) 
we define $N^2, N\otimes N$ matrices $e^{ij}$, i,j=1,2,..,N whose 
matrix elements are given by ${e^{ij}}_{kl} 
\equiv \delta^{i}_k \delta^{j}_{l}$. The generators belonging to 
the first fundamental representation can now be written as: 
\bea 
H^{a}(1) = \sum_{i=1}^{a} e^{ii}-a e^{a+1a+1}, ~~~ 
E^{\alpha}(1) = e^{ij}, ~~ {i > j} 
\label{gen}
\eea
\noindent The higher fundamental representations can be built 
out of  (\ref{gen}).  Following the previous section on SU(3), we 
introduce (N-1) sets of harmonic 
oscillators creation-annihilation operators $(a_{i}(F), a^{\dagger}
_{i}(F)), F = 1, 2,.., N-1; i=1,2,...{}^{N}C_{F}$. 
We now have the Hilbert space characterized by the 
occupation  numbers $n_{i}(F)$ corresponding to each of these 
oscillators. 
At this stage, it is convenient to think of $a_{i}(F)$ and 
$n_{i}(F)$ as the $i^{th}$ components of ${}^{N}C_{F}$  
dimensional vectors $\vec{a}(F)$ and $\vec{n}(F)$ 
respectively.  An arbitrary occupation numbers state can now 
be denoted by $\vert \vec{n}(1),....,\vec{n}(N - 1)) \rangle$ 
where $\vec{n}({F})$
is a set containing ${}^{N}C_{F}$ integers $n_{i}(F), 
(i=1,2,...{}^{N}C_{F})$.  We  define the SU(N) charge 
operators as: 
\bea
{\cal{Q}}_{a} \equiv \sum_{F=1}^{N-1}~ 
\sum_{i,j=1}^{{}{{}^{N}}C_{F}}
a^{\dagger}_{i}(F) (H^{a}(F))_{ij} a_{j}(F) \nonumber \\
{\cal{E}}^{ij} \vert_{i\neq j} \equiv \sum_{F=1}^{N-1}~
\sum_{k,l=1}^{{}{{}^{N}}C_{F}} 
a^{\dagger}_{k}(F) (E^{ij}(F))_{kl} a_{l}(F) 
\label{sunch} 
\eea 
It is easy to check that the operators defined in (\ref{sunch}) satisfy 
the SU(N) Lie algebra (\ref{cartan}). Just like in the case of SU(2) and 
SU(3), the (N - 1) Casimirs of SU(N) are simply the (N - 1) types of number 
operators, i.e,: 
\bea
{\cal C}(F) \equiv \sum_{i=1}^{{}^{N}C_{F}}~a^{\dagger}_{i}(F)a_{i}(F), ~~~~ F=1,2,..,(N - 1). 
\label{casn} 
\eea 
We now define (N-1) complex 
vectors\footnote{Note that for N=3, we 
get the results of section 2 with $\vec{z}(1) \equiv \vec{z}$ and 
$\vec{z}(2) \equiv \vec{w}$.} $\vec{z}(F)$ of dimensions 
${}^{N}C_{F}$ respectively. The SU(N) coherent states which are eigenstates 
of the (N-1) Casimirs are defined as \cite{mmn}:  
\beq
|\vec{z}(1),...,\vec{z}(N-1)>_{(c_1,..,c_{N-1})} ~\equiv ~ 
\Big[\prod_{F=1}^{N-1}
\delta_{({\cal C}[F],c_{F})}\Big]~
\exp ~\sum_{F=1}^{N-1}~(\vec{z}(F) \cdot \vec{a}^\dagger(F)) 
|\vec{0},...,\vec{0}> 
\label{csfn} 
\eeq
In (\ref{csfn}), the (N - 1) $\delta$ function constraints imply that the 
states on the left satisfy: 
\bea 
{\cal C}[F] \vert \vec{z}(1),..,\vec{z}(N-1) \rangle_{(c_{1},..,c_{N-1})} 
& = & c_F~ \vert \vec{z}(1),..,\vec{z}(N-1) \rangle_{(c_1,..,c_{N-1})}.
\label{evenn} 
\eea 
The coherent states in (\ref{csfn}) are defined on the SU(N) manifold 
\cite{mmn}.  On the other hand, the fixed SU(N) charge coherent states 
are given by:
\bea
\vert \vec{z}(1),..,\vec{z}(N-1) \rangle_{q_1,...q_{N-1}}
\equiv \Big[\prod_{a=1}^{(N-1)} \delta_{[{\cal{Q}}_a,~
q_{a}]}\Big]~ 
exp \sum_{F=1}^{N-1} \vec{z}(F).\vec{a}^{\dagger}(F) 
\vert \vec{0},...,\vec{0} \rangle 
\label{dee} 
\eea  
\noindent Again, (\ref{dee}) is a generalization of (\ref{cs3}) and 
(\ref{csf}).  The coherent states (\ref{dee}) are defined 
on the complex manifold ${(C)}^{2[2^{(N-1)}-1]}$.  
At this stage, for the sake of simplicity, we consider only 
F=1 and F = (N - 1) representations (i.e,  $\vec{n}(F) = 0$ 
for F = 2, 3,...(N-2)). They are both N dimensional 
and conjugate to each other. In this simple case, we further denote 
$\vec{n}(F=1) \equiv \vec{n}$ and $\vec{n}(F=N-1) \equiv \vec{m}$. The 
(N - 1) $\delta$ function constraints 
in (\ref{dee}) demand that the  occupation number 
be the the eigenvectors of the (N - 1) charge operators ${\cal{Q}}^{a}$ 
given in  (\ref{sunch}): 

\bea 
{\cal Q}_{a}~ \vert {}^{{n}_1,~{n}_2,..., {n}_N}_{m_1, m_2,..., m_N}\rangle
= q_{a} ~\vert {}^{{n}_1,~{n}_2,..., {n}_N}_{m_1, m_2,..., m_N}\rangle, 
~~ F=1, 2,...,(N - 1).  
\label{nevn} 
\eea

\vspace{0.2cm}
\noindent The (N - 1) equations in (\ref{nevn}) can be easily solved 
for the first (N - 1) occupation numbers $(n_1, n_2,...,n_{(N - 1)})$ 
in terms of the (N - 1) SU(N) charges $(q_1, q_{2},...q_{(N - 1)})$. 
\bea 
n_{i} = n_{N} + l_i + \sum_{a=i}^{N-1} a (m_a-m_{a+1}), ~~~ 
i=1, 2,.., N-1.  
\label{cfn}
\eea  
\noindent where $l_{i} \equiv \sum_{a=i}^{N-1} 
\Big(\frac{1}{a}(q_a-q_{a-1})$. The equation (\ref{cfn}) is 
the SU(N) generalized form of the equation (\ref{cf}) for SU(3).  
We can now directly write down the SU(N) analog of (\ref{exp44}): 
\bea
|\vec{z},\vec{w}>_{(q_1,..,q_{N-1})} 
&=& 
exp~\Big[\frac{z_1..z_{N}}{N_1..N_{N-1}} 
~a_1^{\dagger}..a^{\dagger}_{N} + \frac{w_1..w_{N}}{M_1..M_{N-1}}
b_1^{\dagger}..b^{\dagger}_{N}\Big] 
\sum_{\nu_1,..,\nu_{N-1} = 0}^{\infty}
C^{0}_{\nu_1,..,\nu_{N-1},0}
\nonumber \\ 
&& \vert {}^{(l_1+\nu_1) ~~ (l_2+\nu_2) .... (l_{N-1} +\nu_{N-1}) ~~0} 
_{~~ \nu_1~~~~~~~~~\nu_2~~~~....~~\nu_{N-1} ~~~~~\phantom{.}0}~\rangle
%_{(q_1,q_2,...,q_{N-1})}   
\label{exp444}
\eea 
In (\ref{exp444}), 
\bea 
C^{0}_{\nu_1,..,\nu_{N-1},0} \equiv \frac{z_{1}^{l_1+\nu_1}
...z_{N-1}^{l_{N-1}+\nu_{N-1}}~w_1^{\nu_1} ...w_{N-1}^{\nu_{N-1}}} 
{{\sqrt{(l_1+\nu_1)!
...(l_{N-1}+\nu_{N-1})!\nu_1!...\nu_{N-1}!}}} 
\label{deff} 
\eea
Like in the SU(3) case, the compact exponential form is: 
\bea
|\vec{z},\vec{w}>_{(q_1,..,q_{N-1})}  =  C_0 
exp\Big[{z_1..z_{N-1}}~\frac{1}{\hat{N}_1..\hat{N}_{N-1}}
a_1^{\dagger}..a_{N-1}^{\dagger}+{w_1..w_{N-1}} 
\frac{1}{\hat{M}_1..\hat{M}_{N-1}}
b_1^{\dagger}..b_{N-1}^{\dagger}\Big] 
\nonumber \\
exp~\Big[z_1w_1\frac{1}{\hat{N}_1}a_1^{\dagger}b_1^{\dagger} 
..+ z_{N-1}w_{N-1}\frac{1}{\hat{N}_{N-1}}a_{N-1}^{\dagger}
b_{N-1}^{\dagger}\Big]  
\vert {}^{l_1 ~~~ l_2 ~~.... l_{N-1} ~~~0} 
_{ 0~~~~0~~~....0 ~~~~~\phantom{..}0}~\rangle    
\hspace{1.5cm} 
\label{exp98} 
\eea
where, 
\bea 
C_0 \equiv \frac{(z_1)^{l_1}}{\sqrt{l_1!}} 
.....\frac{(z_{N-1})^{l_{N-1}}}{\sqrt{l_{N-1}!}} 
\label{exp100} 
\eea 
The coherent states with SU(N) charges in (\ref{exp444}) satisfy: 
\bea 
{\cal{H}}^{a} ~~\vert \vec{z},\vec{w} 
\rangle_{q_1,q_2,...q_{N-1}} & = & q_{a} ~~
\vert \vec{z},\vec{w} \rangle_{q_1,q_2,...q_{N-1}}, a =1,2...,N - 1 \nonumber 
 \\ 
a_{i}b_{i}  \vert \vec{z},\vec{w} 
\rangle_{q_1,q_2,...q_{N-1}} & = & z_{i}w_{i}
\vert \vec{z},\vec{w} \rangle_{q_1,q_2,...q_{N-1}} 
 ~~ i=1,2,...,N 
\nonumber \\  
\label{evennn}
\Big(\prod_{i=1}^{N} a_{i}\Big)
\vert \vec{z},\vec{w} \rangle_{q_1,q_2,...q_{N-1}}
& = & \Big(\prod_{i=1}^{N} z_{i}\Big) 
\vert \vec{z},\vec{w} \rangle_{q_1,q_2,...q_{N-1}} \\
\Big(\prod_{i=1}^{N} b_{i}\Big) 
\vert \vec{z},\vec{w} \rangle_{q_1,q_2,...q_{N-1}}
& = & \Big(\prod_{i=1}^{N} w_{i}\Big) 
\vert \vec{z},\vec{w} \rangle_{q_1,q_2,...q_{N-1}}
\nonumber 
\eea 
The equations in (\ref{evennn}) are the SU(N) analog of (\ref{eve2}) 
for SU(3). 

\subsection{Nonlinear Formulation} 

As in the case of SU(3), we now define the non-linear SU(N) 
coherent states by generalizing (\ref{nl}). We now have (N - 1) 
conditions: 
\bea 
\Big[f_F(\hat{N}_1,...,\hat{N}_{N}) \prod_{i=1}^{{}^{N}C_{F}} a_{i}(F)\Big]
~ |\vec{z}(1)..,\vec{z}(N-1);f_1,..,f_F> \hspace{5cm} \nonumber \\
\equiv \Big[\prod_{i=1}^{{}^{N}C_{F}}~z_{i}(F)\Big]~ 
|\vec{z}(1)..,\vec{z}(N-1);f_1,..,f_F>, ~~F=1,2..,N-1
\label{nln} 
\eea 
For the sake of simplicity, we again restrict ourselves to only  F = 1 and 
F = (N - 1) fundamental representations. Denoting $f_{F=1}$ and $f_{F=N-1}$ 
by f and g respectively, the equations in (\ref{nln}) reduce to:  
\bea 
f(\hat{N}_1,...,\hat{N}_N) a_1..a_N \vert\vec{z},\vec{w},f,g \rangle_{q_1,..,q_{N-1}} 
& = & z_1...z_N \vert\vec{z},\vec{w},f,g \rangle_{q_1,..,q_{N-1}} \nonumber \\
g(\hat{M}_1,...,\hat{M}_N) b_1..b_N \vert\vec{z},\vec{w},f,g \rangle_{q_1,..,q_{N-1}}  
& = & w_1...w_N \vert\vec{z},\vec{w},f,g \rangle_{q_1,..,q_{N-1}}. 
\label{nlnn}
\eea 
Again, the construction of $\vert\vec{z},\vec{w},f,g \rangle_{q_1,..,q_{N-1}}$ 
is  similar to that of section 3 for SU(3). We define: 
\bea
\hat{H}_{f}(\hat{N}_1,\hat{N}_2,...,\hat{N}_{N-1}) &\equiv& \frac{1}{\hat{N}_1 
\hat{N}_2..\hat{N}_{N-1} ~f(\hat{N}_1-1,\hat{N}_2-1,...,\hat{N}_{N-1}-1)} 
\nonumber \\
\hat{H}_{g}(\hat{M}_1,\hat{M}_2,...,\hat{M}_{N-1}) &\equiv& \frac{1}{\hat{M}_1 
\hat{M}_2..\hat{M}_{N-1}~g(\hat{M}_1-1,\hat{M}_2-1,...,\hat{M}_{N-1}-1)}. 
\label{idi} 
\eea 
Like in the SU(2) and SU(3) cases we  now get: 
\bea 
|\vec{z},\vec{w},f,g>_{q_1,..,q_{N-1}} ~~ = ~~  
exp~\Big[{z_1..z_N}~{\hat{H}_f}~a_1^{\dagger}..a_N^{\dagger}
+{w_1..w_N}~{\hat{H}_g}~ b_1^{\dagger}..b_{N}^{\dagger}\Big] \nonumber \\ 
\sum_{\nu_1,..,\nu_{N-1}= 0}^{\infty}
C^{0}_{\nu_1,..,\nu_{N-1},0}~~ 
\vert {}^{(l_1+\nu_1) ~~ (l_2+\nu_2) .... (l_{N-1} +\nu_{N-1}) ~~0} 
_{~~ \nu_1~~~~~~~~~\nu_2~~~~....~~\nu_{N-1} ~~~~~\phantom{.}0}~\rangle 
\nonumber \\
= 
C_0~exp~\Big[{z_1..z_N}~{\hat{H}_f}~a_1^{\dagger}..a_N^{\dagger}
+{w_1..w_N}~{\hat{H}_g}~ b_1^{\dagger}..b_{N}^{\dagger}\Big] \nonumber \\ 
exp~\Big[z_1 w_1 \frac{1}{\hat{N}_1} a_{1}^{\dagger} b_{1}^{\dagger}+ ...
+ z_{N-1}w_{N-1} \frac{1}{\hat{N}_{N-1}} a_{N-1}^{\dagger} 
b_{N-1}^{\dagger}\Big] 
\vert {}^{l_1 ~~~ l_2 ~~.... l_{N-1} ~~~0} 
_{ 0~~~~0~~~....0 ~~~~~\phantom{..}0}~\rangle 
\label{exp77} 
\eea 
In (\ref{exp77}) $C^{0}_{\nu_1,..,\nu_{N-1},0}$ and $C_0$ are given by 
(\ref{deff}) and (\ref{exp100}) respectively. We also note that 
$|\vec{z},\vec{w},f=1,g=1>_{q_1,..,q_{N-1}} \equiv 
|\vec{z},\vec{w}>_{q_1,..,q_{N-1}}$.  
All the explicit constructions in this section can be generalized to 
include the other SU(N) fundamental representations.  

\vspace{0.8cm} 

\section{\bf Summary and Discussion} 

In this paper we have exploited the representation of  $SU(N)$ 
Lie algebra in terms of harmonic oscillator creation and annihilation 
operators to construct SU(N) coherent states with fixed (N - 1) 
charges belonging to the SU(N) Cartan subspace . These fixed 
charge coherent states are obtained by simply modifying the constraints 
for the corresponding generalized coherent states which 
are defined over different possible irreducible representations of the 
group.  Further, it is easy 
to generalize this Schwinger boson construction to define non-linear 
SU(N) coherent states.  We should also mention that the techniques 
described in this work using harmonic 
oscillators are extremely simple compared to the standard method. This 
is because the standard definition of SU(N) coherent states belonging to 
a representation R of SU(N) is: 
\bea 
|g>_{R} \equiv T_{R}(g) |\mu>_{R}   
\label{dis} 
\eea
where g is an SU(N) group element characterized by $(N^2 - 1)$ compact 
parameters, $T_{R}(g)$ is the corresponding matrix and $|\mu>_{R}$ is 
an arbitrary weight vector in the representation R respectively. Therefore, 
to define $|g>_{R}$, we not only need to know the group manifold with the 
Haar measure, we also need to know all the representations 
of SU(N). Further, for bigger SU(N) groups and their higher representations 
it becomes extremely difficult to analyze (\ref{dis}) because of large matrix 
structures and too many parameters describing the group manifolds. This 
should be contrasted with (\ref{cs2}), (\ref{csc}) and (\ref{csfn}) where all 
these inputs are part of the formulation itself. Infact, this leads 
to enormous simplifications for $N \ge 3$.  We,therefore, expect that 
generalization of our techniques to an arbitrary Lie group will be 
convenient and useful. The work in this direction is in progress and 
will be reported elsewhere.

\vspace{0.8cm}

\noindent{\bf Acknowledgment} 

\vspace{0.5cm}

\noindent One of the authors (MM) thanks Prof. H. S. Mani for reading the manuscipt
carefully.

\end{document}